\documentclass[
 reprint,
superscriptaddress,
 amsmath,
 amssymb,
 prl,
 longbibliography,
 color,
 tabulary,
 floatfix
]{revtex4-1}

\usepackage{blindtext}
\usepackage{graphicx}
\usepackage{natbib}
\usepackage{siunitx}
\usepackage{textcomp}
\usepackage{gensymb}
\usepackage{multirow}
\usepackage[dvipsnames]{xcolor}
\usepackage{hyperref}
\hypersetup{colorlinks=true,linkcolor=blue,filecolor=magenta,urlcolor=cyan}
\usepackage[nameinlink, noabbrev]{cleveref}
\usepackage{soul}

\newcommand{\beq}{\begin{equation}}
\newcommand{\eeq}{\end{equation}}

\begin{document}
	\title{Nonlinear Dynamics of Preheating after Multifield Inflation \\ with Nonminimal Couplings}

\newcommand{\Kenyon}{Department of Physics, Kenyon College, Gambier, Ohio 43022, USA}
\newcommand{\Nikhef}{Nikhef, Science Park 105, 1098XG Amsterdam, The Netherlands}
\newcommand{\Lorentz}{Lorentz Institute for Theoretical Physics, Leiden University, 2333CA Leiden, The Netherlands}
\newcommand{\MIT}{Department of Physics, Massachusetts Institute of Technology, Cambridge, Massachusetts 02139, USA}
\newcommand{\Case}{CERCA/ISO, Department of Physics, Case Western Reserve University, 10900 Euclid Avenue, Cleveland, OH 44106}

\author{Rachel~Nguyen}
\email{nguyenr@kenyon.edu}
\affiliation{\Kenyon}
\author{Jorinde~van~de~Vis}
\email{jorindev@nikhef.nl}
\affiliation{\Nikhef}
\author{Evangelos~I.~Sfakianakis}
\email{evans@nikhef.nl}
\affiliation{\Nikhef} 
\affiliation{\Lorentz}
\author{John~T.~Giblin,~Jr.}
\email{giblinj@kenyon.edu}
\affiliation{\Kenyon}
\affiliation{\Case}
\author{David~I.~Kaiser}
\email{dikaiser@mit.edu}
\affiliation{\MIT}

	\begin{abstract}
	We study the post-inflation dynamics of multifield models involving nonminimal couplings using lattice simulations to capture significant nonlinear effects like backreaction and rescattering. We measure the effective equation of state and typical time-scales for the onset of thermalization, which could affect the usual mapping between predictions for primordial perturbation spectra and measurements of anisotropies in the cosmic microwave background radiation. For large values of the nonminimal coupling constants, we find efficient particle production that gives rise to nearly instantaneous preheating. Moreover, the strong single-field attractor behavior that was previously identified persists until the end of preheating, thereby suppressing typical signatures of multifield models.  We therefore find that predictions for primordial observables in this class of models retain a close match to the latest observations. 
	\end{abstract}
	
	\maketitle

{\it Introduction}. Post-inflation reheating plays a critical role in our understanding of the very early Universe (see Ref.~\cite{AHKK} for a recent review). By the end of the reheating phase --- and before big-bang nucleosynthesis (BBN) can commence \cite{BBNreview} --- the Universe must achieve a radiation-dominated equation of state and become filled with (at least) a thermal bath of Standard Model particles at an appropriately high temperature. Although the earliest stages of reheating can be studied within a linearized approximation, some of the most critical processes arise from nonlinear physics, including backreaction and rescattering among the produced particles.

In addition to setting appropriate conditions for BBN, the reheating phase plays a critical role in comparisons between inflationary predictions and recent high-precision measurements of the cosmic microwave background (CMB). In particular, if there were a prolonged period after inflation before the Universe attained a radiation-dominated equation of state (EOS), that would impact the mapping between perturbations on observationally relevant length-scales and when those scales first crossed outside the Hubble radius during inflation \cite{AdsheadEastherReheat,DaiKamionkowski,VenninWandsCurv,Lozanov:2016hid}. Residual uncertainty on the duration of reheating, $N_{\rm reh}$, is now comparable to statistical uncertainties in measurements of CMB spectral observables. Hence understanding the time-scale $N_{\rm reh}$ is critical for evaluating observable predictions from inflationary models. 

In this Letter we study the nonlinear dynamics of the early preheating phase of reheating in a well-motivated class of models. These models include multiple scalar fields, as typically found in realistic models of high-energy physics \cite{Mazumdar:2010sa,baumann+mcallister_15}; and each scalar field, $\phi$, has a nonminimal coupling to the spacetime Ricci curvature scalar, $R$, of the form $\xi \phi^2 R$.  Such nonminimal couplings are quite generic: they are induced by quantum corrections for any self-interacting scalar field in curved spacetime, and they are required for renormalization \cite{Callan:1970ze,Bunch:1980br}. Moreover, the dimensionless coupling constants, $\xi$, grow with energy-scale under renormalization-group flow, with no UV fixed point \cite{Odintsov:1990mt}. Hence they can attain large values at inflationary energy scales. Upon transforming to the Einstein frame, such models feature curved field-space manifolds \cite{DKconformal}. 

Multifield models with nonminimal couplings naturally yield a plateau-like phase of inflation at large field values, of the sort most favored by recent observations \cite{Planck2018Inflation}. During inflation the fields generically evolve within a single-field attractor, thereby suppressing typical multifield effects that could spoil agreement with observations, such as large primordial non-Gaussianities and isocurvature perturbations \cite{KMSbispectrum,KSprl,SSKisocurv}.

Previous work, which studied the onset of preheating in this class of models semi-analytically, identified three regimes that yielded qualitatively distinct behavior: $\xi \lesssim {\cal O} (1)$, $\sim {\cal O} (10)$, and $\gtrsim {\cal O} (10^2)$ \cite{Preh1,Preh2,Preh3}. In this Letter we significantly expand this work, employing lattice simulations to study the complete preheating phase, deep into the nonlinear regime. We restrict attention to coupled scalar fields, and neglect the production of Standard Model particles such as fermions or gauge fields \cite{Greene:1998nh,Greene:2000ew,Peloso:2000hy,Tsujikawa:2000ik,Davis:2000zp,GarciaBellido:2003wd,Bezrukov:2008ut,GarciaBellido:2008ab,Dufaux:2010cf,Allahverdi:2011aj,Deskins:2013lfx,Adshead:2015kza,Adshead:2017xll,Sfakianakis:2018lzf}. Nonetheless, we are able to analyze the typical time-scales required for the Universe to achieve a radiation-dominated EOS; for the produced particles to backreact on the inflaton condensate, ultimately draining away its energy; and for rescattering among the particles to yield a thermal spectrum.  For large couplings, $\xi \gtrsim 10^2$, of the sort encountered in Higgs inflation \cite{HiggsInflation}, we find very efficient preheating, typically completing within the first two $e$-folds after the end of inflation, thereby protecting the close match between predictions for primordial observables and the latest CMB measurements.

{\it Model}. In the Jordan frame, the nonminimal coupling between the $N$ scalar fields and the spacetime Ricci scalar $\tilde{R}$ remains explicit in the action through the term $f (\phi^I) \tilde{R}$. Upon rescaling $\tilde{g}_{\mu\nu} (x) \rightarrow g_{\mu\nu} (x) = \Omega^2 (x) \tilde{g}_{\mu\nu} (x)$, with $\Omega^2 = 2 f(\phi^I ) / M_{\rm pl}^2$, we transform the action into the Einstein frame. (Here $M_{\rm pl} \equiv 1/ \sqrt{ 8 \pi G} = 2.43 \times 10^{18}$ GeV is the reduced Planck mass.) The Einstein-frame potential is stretched by the conformal factor, $V (\phi^I) = \tilde{V} (\phi^I)/ \Omega^4$, compared to the Jordan-frame potential $\tilde{V} (\phi^I)$. Taking canonical scalar fields in the Jordan frame, the nonminimal couplings induce a curved field-space manifold in the Einstein frame, with field-space metric given by ${\cal G}_{IJ} (\phi^K) = [ M_{\rm pl}^2 / (2f) ] \{ \delta_{IJ} + 3 f_{, I} f_{, J} / f \}$ \cite{DKconformal}. The equation of motion for the fields in the Einstein frame is then
\beq
\Box \phi^I + g^{\mu\nu} \Gamma^I_{\> JK} \partial_\mu \phi^J \partial_\nu \phi^K - {\cal G}^{IJ} V_{, J} = 0 ,
\label{eomphi}
\eeq
where $\Gamma^I_{\> JK} (\phi^L )$ is the Christoffel symbol constructed from ${\cal G}_{IJ}$. We consider an unperturbed, spatially flat Friedmann-Lema\^{i}tre-Robertson-Walker (FLRW) spacetime metric, so the Einstein field equations yield $H^2 (t) = \rho_{\rm total} / (3 M_{\rm pl}^2 )$, where $\rho_{\rm total}$ is the total energy density of the system, $H (t) \equiv \dot{a} / a$, and overdots denote derivatives with respect to cosmic time.

We consider two-field models, $\phi^I = \{ \phi, \chi \}$, with 
%%%%
\beq
\begin{split}
    f (\phi^I) &= \frac{1}{2} \left[ M_{\rm pl}^2 + \xi_\phi \phi^2 + \xi_\chi \chi^2 \right] , \\
    \tilde{V} (\phi^I) &= \frac{\lambda_\phi }{4} \phi^4 + \frac{g}{2} \phi^2 \chi^2 + \frac{ \lambda_\chi}{4} \chi^4 .
\end{split}
\label{fV}
\eeq
%%%%%%
The topography of the Einstein-frame potential generically includes ``ridges'' and ``valleys'' along certain directions $\chi/ \phi = const$.
For non-fine-tuned parameters, the fields quickly fall to a local minimum (valley) of the potential, and the background dynamics obey a strong ``single-field attractor" \cite{KSprl,SSKisocurv,Preh1}. For symmetric couplings, with $\xi_\phi = \xi_\chi$ and $\lambda_\phi = g = \lambda_\chi$,
any initial angular motion within field space damps out within a few $e$-folds after the start of inflation, and the system flows toward the minimum of the potential along a single-field trajectory \cite{GKSHiggs}. Within a single-field attractor, the predictions for the spectral index $n_s$, the tensor-to-scalar ratio $r$, the running $\alpha = d n_s / d \ln k$, primoridal non-Gaussianities $f_{\rm NL}$, and isocurvature perturbations $\beta_{\rm iso}$ remain consistent with the latest observations across large regions of phase space and parameter space \cite{KSprl,SSKisocurv,Preh1}.

Field fluctuations in these models are sensitive to the curvature of the field-space manifold, which is greatest near the origin. During preheating, as the inflaton condensate oscillates through zero, the effective mass for the fluctuations $\delta \chi$ receives quasi-periodic ``spikes" proportional to a component of the field-space Riemann tensor. In the limit $\xi_I \gg 1$, these scale as ${\cal R}^\chi_{\>\> \phi \phi \chi} \propto \xi_\phi$.
These large ``spikes" lead to sharp violations of the adiabatic condition for those modes, driving efficient particle production \cite{Preh1,Preh2,Preh3, Ema:2016dny}. 

Within the single-field attractor, the amplitude of primordial perturbations scales as $[ \lambda_\phi / \xi_\phi^2 ]^{1/2}$ \cite{KSprl}. Present constraints on the tensor-to-scalar ratio therefore require $\lambda_\phi / \xi_\phi^2 \leq 1.4 \times 10^{-8}$. We fix $\lambda_\phi / \xi_\phi^2= 10^{-8}$ and consider various values for $\xi_\chi / \xi_\phi$, $\lambda_\chi / \lambda_\phi$, and $g / \lambda_\phi$. We consider two typical cases: (A) $\xi_\chi = 0.8 \xi_\phi$, $g = \lambda_\phi$, and $\lambda_\chi = 1.25 \lambda_\phi$; and (B) $\xi_\chi = \xi_\phi$, $\lambda_\phi = g = \lambda_\chi$.
For the ``generic'' case (A) the single-field attractor lies along $\chi=0$, while we are free to choose the same attractor direction for the symmetric case (B).
Once the ratios of couplings are fixed, the dynamics of the system change as we vary $\xi_\phi$ across $\lesssim {\cal O} (1), \sim {\cal O} (10)$, and $\gtrsim {\cal O} (10^2)$.

{\it Results}.  
We employ a modified version of {\sc GABE} (Grid and Bubble Evolver) \cite{GABE} to evolve the fields and the background, according to Eq.~(\ref{eomphi}) and the Friedmann equation.  Whereas the original software was used to simulate nonminimally coupled degrees of freedom \cite{Child:2013ria}, we have modified the code significantly to allow for a curved field-space metric in both the dynamics of the fields as well as the initial conditions.  We start the simulations when inflation ends, defined by $\epsilon (t_{\rm init})= 1$ where $\epsilon \equiv - \dot{H} / H^2$; the Hubble scale at this time is $H_{\rm end}$.  We use a grid with $\mathcal{N} = 256^3$ points and a comoving box size $L = \pi/H_{\rm end}$ so that the longest wavelength in our spectra corresponds to $k = H_{\rm end}/2$. We match the two-point correlation functions of $\phi (t_{\rm init},{\bf x})$ and $\chi (t_{\rm init}, {\bf x} )$ to corresponding distributions for quantized field fluctuations. Fourier modes of the quantized fluctuations evolving during inflation within the single-field attractor may be parameterized as 
$\delta \phi_k^I = \sqrt{ {\cal G}^{II }} \,v^I_k (\tau) / a(\tau)$ (no sum on $I$),
where $d\tau \equiv dt / a(t)$ is conformal time \cite{Preh1}. Near the end of inflation, we use the Wentzel-Kramers-Brillouin (WKB) approximation to estimate amplitudes 
$\vert v_k^I (\tau_{\rm init}) \vert = [ 2 \Omega_{(I)} (k, \tau_{\rm init}) ]^{-1/2}$,
where $\Omega_{(I)}^2 (\tau) = k^2 + a^2 (\tau) \, m_{ {\rm eff}, I}^2 (\tau)$. The effective masses $m_{{\rm eff}, I}^2$ include distinct contributions from the curvature of the potential and from the curvature of the field-space manifold, and are analyzed in detail in Refs.~\cite{Preh1,Preh2,Preh3}. (Here we neglect contributions from coupled metric perturbations.)  The initial spectra of the fields are subject to a window function that suppresses high-momentum modes above some UV suppression scale, $k_{\rm UV} = 50\, H_{\rm end}$.

\begin{figure}[t!]
    \centering
    \includegraphics[width=3.4in]{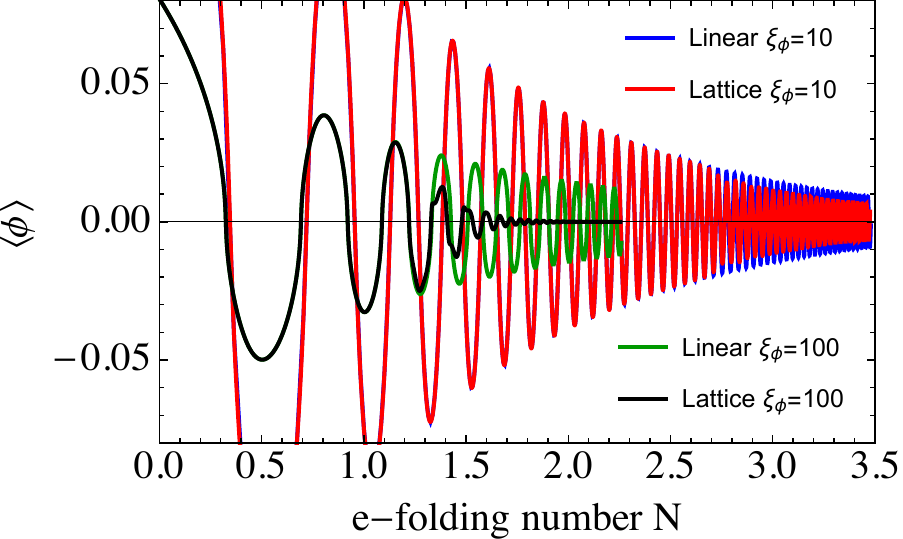} 
    \caption{\small The evolution of the inflaton condensate (in units of $M_{\rm pl}$) versus $e$-folds $N$ after the end of inflation for Case A with $\xi_\phi = 10, 100$, as calculated in linearized analysis (blue, green) and as computed from the spatial average $\langle \phi \rangle$ on the lattice (red, black).}
    \label{fig:xi10background}
\end{figure}

\begin{figure}[t!]
    \centering
    \includegraphics[width=3.4in]{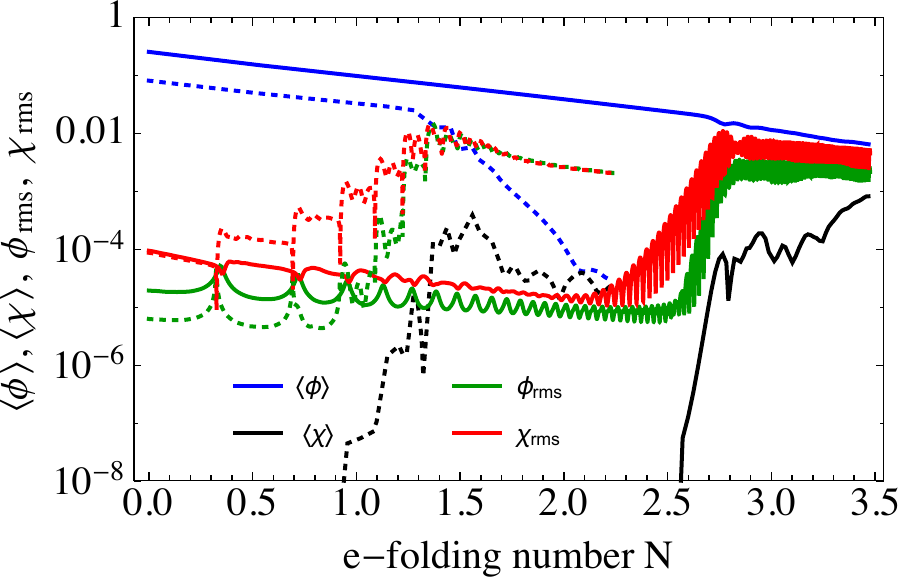}
    \caption{\small Lattice evolution of various fields (in units of $M_{\rm pl}$) versus $e$-folds $N$ after the end of inflation for Case A with $\xi_\phi = 10$ (solid) and $\xi_\phi=100$ (dotted): peak values of the spatial averages $\langle \phi \rangle$ (blue) and $\langle \chi \rangle$ (black); and values of the fluctuations $\phi_{\rm rms}$ (green) and $\chi_{\rm rms}$ (red).}
    \label{fig:xi10var}
\end{figure}

Figs.~\ref{fig:xi10background} and \ref{fig:xi10var} show results for Case A with $\xi_\phi = 10, 100$. In Fig.~\ref{fig:xi10background}, we plot the evolution of the inflaton condensate after the end of inflation as calculated in a linearized treatment (akin to Ref.~\cite{Preh3}), and as calculated from the spatial average $\langle \phi \rangle$ on the lattice. Backreaction of produced particles --- which is absent in linearized analyses --- becomes significant beginning around $2.7$ $e$-folds after the end of inflation for $\xi_\phi=10$. For $\xi_\phi=100$ backreaction is strong enough to completely drain the inflaton condensate within the first $2$ $e$-folds. Fig.~\ref{fig:xi10var} shows the evolution of the peak values of the spatial averages $\langle \phi \rangle$ and $\langle \chi \rangle$ as well as the growth of fluctuations, characterized by $\phi_{\rm rms} \equiv \sqrt{ \langle \phi^2 \rangle - \langle \phi \rangle^2}$ and $\chi_{\rm rms} \equiv \sqrt{ \langle \chi^2 \rangle - \langle \chi \rangle^2}$. (Growth of field fluctuations corresponds to particle production \cite{AHKK}.) We have confirmed that the early growth of $\delta \phi$ and $\delta \chi$ fluctuations in our lattice simulations closely matches the behavior calculated via Floquet analysis in Ref.~\cite{Preh2}. Beginning around 2.6 $e$-folds, nonlinear rescattering among the $\delta \chi$ fluctuations drives rapid growth of the $\delta \phi$ fluctuations for $\xi_\phi=10$. For $\xi_\phi=100$ the same effect occurs within the first $e$-fold. Backreaction and rescattering generally become significant at distinct times as one varies couplings \cite{LatticeFollowUp}.

\begin{figure}[t!]
    \centering
    \includegraphics[width=3.4in]{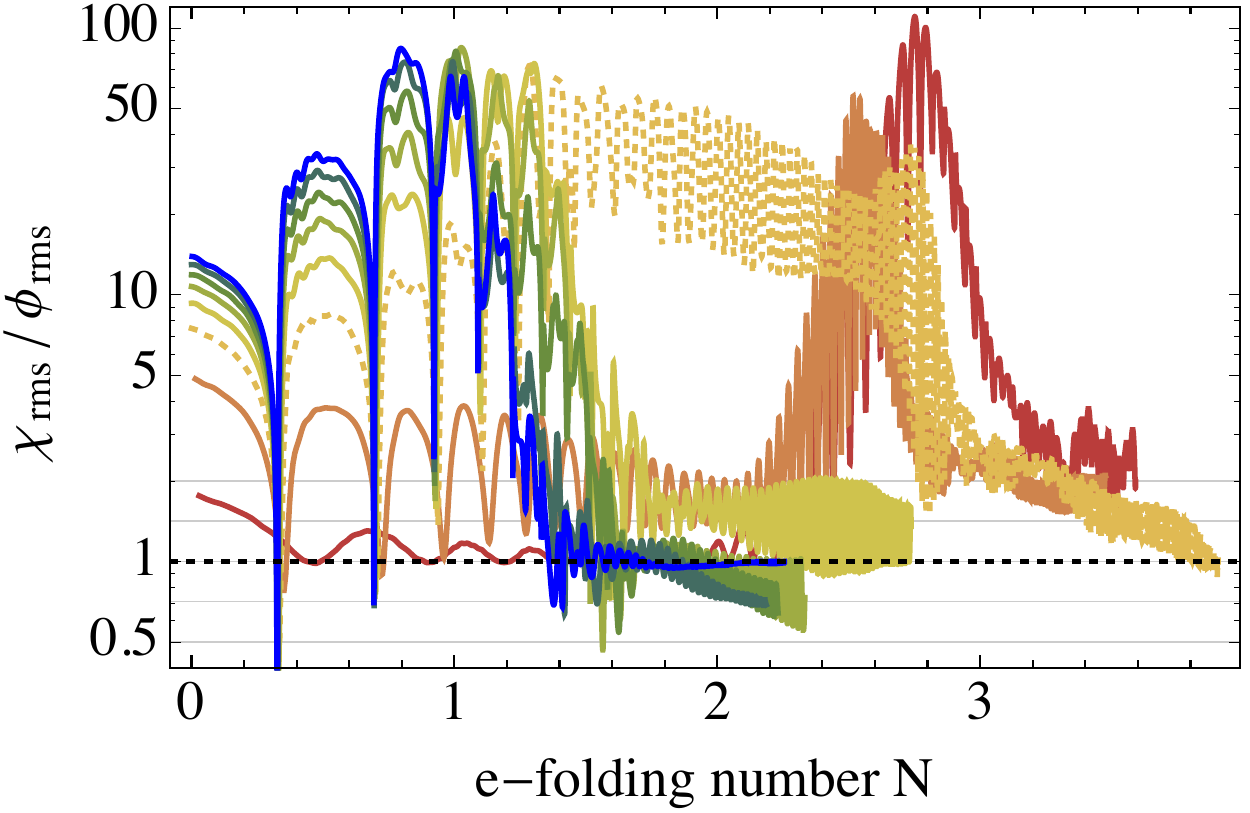}
    \caption{\small The ratio $\chi_{\rm rms} / \phi_{\rm rms}$ versus $e$-folds $N$ after the end of inflation, for Case A with $\xi_\phi = 1, 10, 25, 40, 55, 70, 85, 100$ (red to blue, respectively).}
    \label{fig:rmsvsxi}
\end{figure}

The dynamics of the $\delta \phi$ and $\delta \chi$ fluctuations vary with coupling $\xi_\phi$, as shown in Fig.~\ref{fig:rmsvsxi}.
For $\xi_\phi=1, 10$ parametric resonance due to the contribution from the potential to $m_{{\rm eff},\chi}^2$ leads to a slow growth of $\delta\chi$ fluctuations; these eventually rescatter, leading to the growth  of $\delta \phi$ fluctuations and lowering the $\chi_{\rm rms} / \phi_{\rm rms}$ ratio.
For $\xi_\phi \ge 40$ the ``Ricci spike" \cite{Preh1,Ema:2016dny} leads to a fast growth of $\delta \chi$ fluctuations. This is seen in Fig.~\ref{fig:rmsvsxi} as an early rise of the $\chi_{\rm rms} / \phi_{\rm rms}$ ratio. When $\chi_{\rm rms}$ grows enough it rescatters  with $\delta \phi$ fluctuations, eventually leading to $\chi_{\rm rms} / \phi_{\rm rms}\sim 1$. The case of $\xi_\phi=25$ is the most interesting, since it displays several distinct phases. The initial growth occurs due to adiabaticity violation caused by the Ricci spike. After $1.5$ $e$-folds the height of the Ricci spike has redshifted, making it comparable to the potential contribution to the effective mass, thereby shutting off particle production \cite{Preh1}. When the Ricci spike redshifts even more, around $2.5$ $e$-folds, a second stage of parametric resonance commences, due to the potential term alone. Subsequently, rescattering enhances the $\delta \phi$ fluctuations,  lowering the $\chi_{\rm rms} / \phi_{\rm rms}$ ratio. The situation is qualitatively similar for the symmetric case (B) \cite{LatticeFollowUp}.

\begin{figure}[t!]
    \centering
    \includegraphics[width=3.4in]{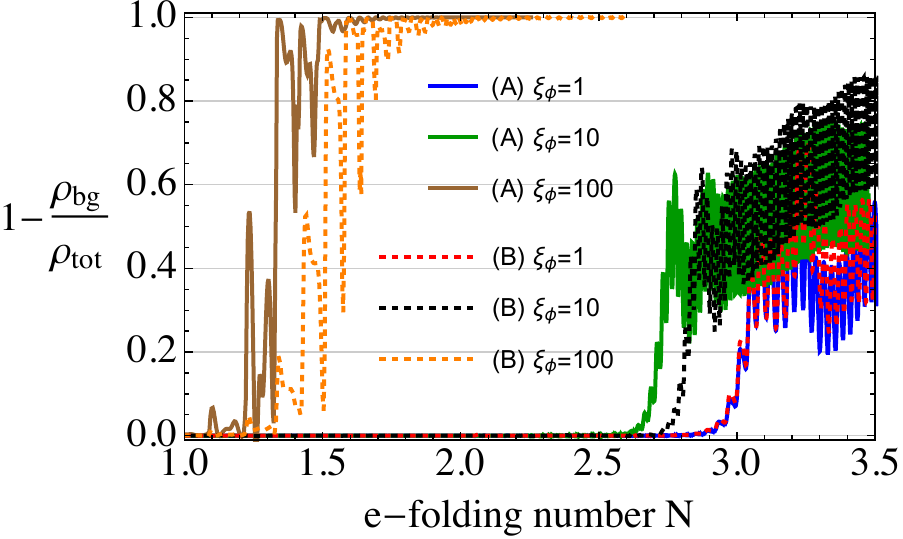}
    \caption{\small The fraction of energy density that has left the inflaton condensate versus $e$-folds $N$ after the end of inflation for the generic case (A) and the symmetric case (B) with $\xi_\phi = 1, 10, 100$. }
    \label{fig:rhoratio}
\end{figure}

The rapid growth of fluctuations yields an efficient transfer of energy from the inflaton condensate into radiative degrees of freedom. Within the single-field attractor, we may approximate the energy density in the inflaton condensate as \cite{Preh1}
%%%%%
\beq
%\rho_{\langle \phi \rangle} 
\rho_{\rm bg} \simeq \frac{1}{2} {\cal G}_{\phi \phi} \langle \dot{\phi} \rangle^2 + \frac{ \lambda_\phi M_{\rm pl}^4 \langle \phi \rangle^4 }{4 ( M_{\rm pl}^2 + \xi_\phi\langle \phi \rangle^2 )^2}  ,
\label{rhocondensate}
\eeq
where we evaluate ${\cal G}_{\phi \phi}$ with $\phi \rightarrow \langle \phi \rangle$ and $\chi \sim 0$. Fig.~\ref{fig:rhoratio} shows that across Cases A and B the fraction of energy density in the inflaton condensate falls sharply within the first few $e$-folds after the end of inflation; for $\xi_\phi \geq 100$, virtually all of the energy density has been transferred out of the inflaton condensate within the first $N = 1.5$ $e$-folds. 

\begin{figure}[t!]
    \centering
    \includegraphics[width=3.4in]{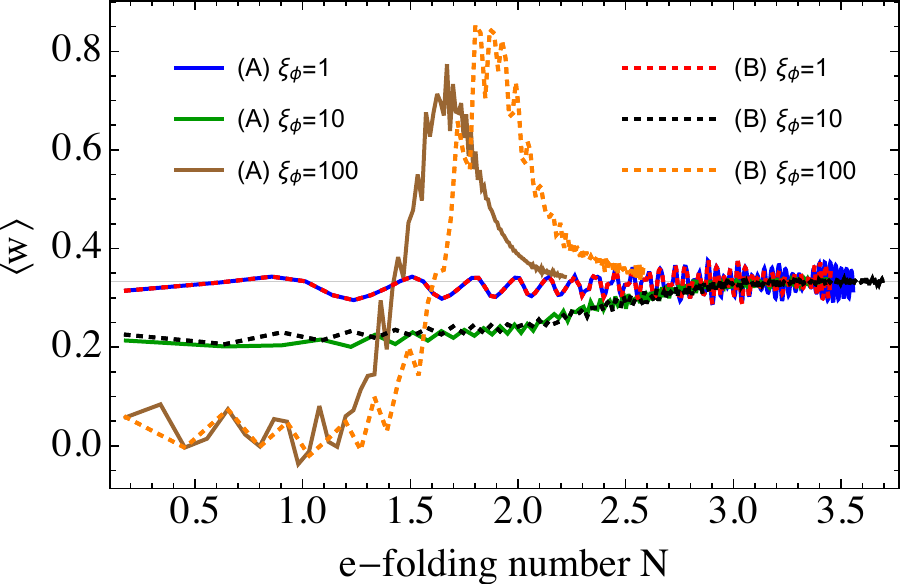}
    \caption{\small The averaged effective equation of state $\langle w \rangle $ for $\xi_\phi= 1, 10, 100$ and the two representative cases, ``generic'' (A) and symmetric (B).  } 
    \label{fig:wvsN}
\end{figure}

The rapid transfer of energy to radiative degrees of freedom is similarly reflected in Fig.~\ref{fig:wvsN}, which shows the evolution of the EOS, $w = p_{\rm total} / \rho_{\rm total}$, where $\rho_{\rm total}$ and $p_{\rm total}$ are the total energy density and pressure for the system, respectively. In this case, the system approaches $w = 1/3$ rapidly for small couplings $\xi_\phi \sim {\cal O} (1)$, because in that regime the Einstein-frame potential for the inflaton approximates a quartic form, so that even the condensate's oscillations correspond to $w \simeq 1/3$ \cite{Preh1}. As $\xi_\phi$ increases, the Einstein-frame potential for $\phi$ approaches a quadratic form, for which the condensate's oscillations behave like $w \simeq 0$ \cite{Preh1}; but in that case, the stronger coupling yields more efficient particle production, so that the system eventually becomes dominated by radiative degrees of freedom. For $\xi_\phi = 100$, we find a transient phase with a stiff EOS, $w > 1/3$, which likely arises because typical momenta for the fluctuations are comparable to $m_{{\rm eff}, I}$, and the contributions to $\rho_{\rm total}$ and $p_{\rm total}$ from kinetic and spatial-gradient terms are weighted by components of ${\cal G}_{IJ}$, which are significant for $\xi_\phi \gg 1$. At later times, as $m_{{\rm eff}, I} \rightarrow 0$, the system relaxes to a gas of massless particles with $w = 1/3$.  Across a wide range of couplings for this family of models, we therefore find that the Universe rapidly achieves a radiation-dominated EOS within $N_{\rm rad} \sim 2 - 2.5$ $e$-folds after the end of inflation. Preheating in $\alpha$-attractor models with $\alpha = {\cal O} (1)$, in contrast, can lead to a prolonged period with $w \simeq 0$ \cite{Iarygina}, shifting the pivot-scale accordingly and thereby offering a means to empirically distinguish between such models and the family we consider here. 

\begin{figure}[t!]
    \centering
    \includegraphics[width=3.4in]{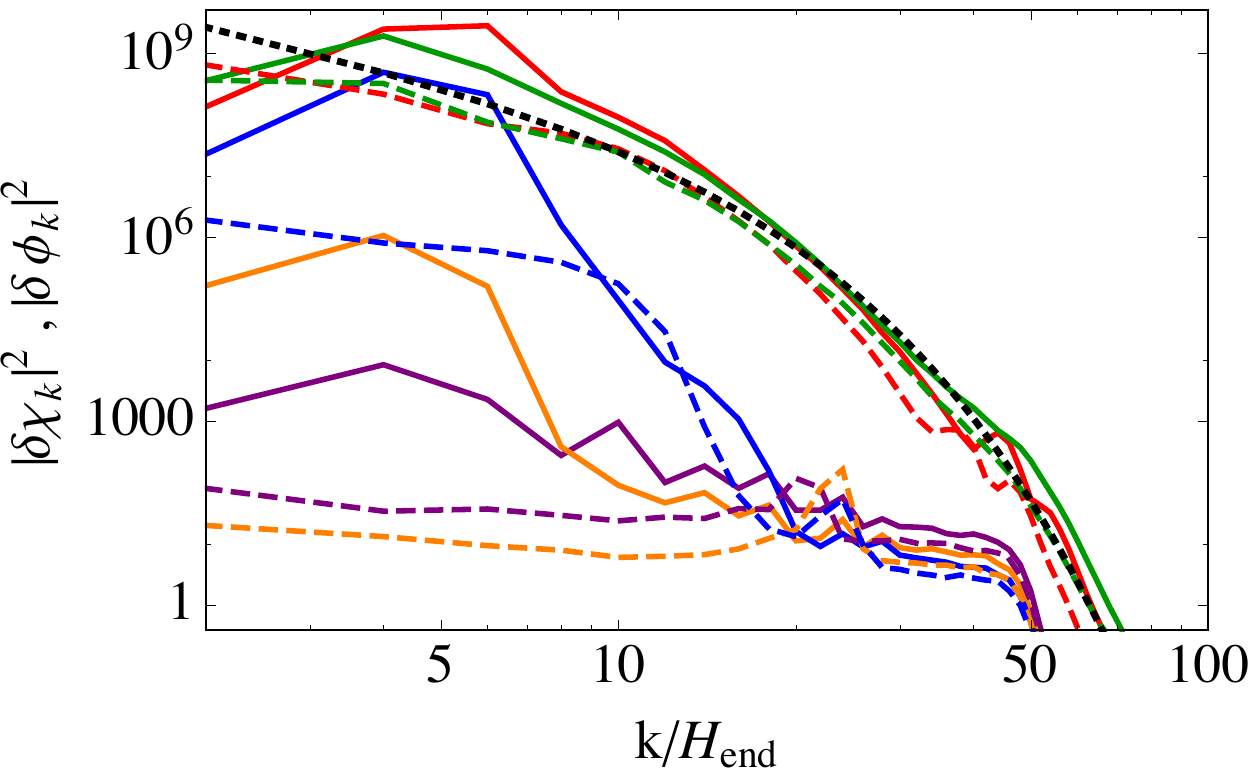}
    \caption{\small  Spectra for the fluctuations $\delta \phi$ (dashed) and $\delta \chi$ (solid) versus $k / H_{\rm end}$, where $k$ is comoving wavenumber, for Case A with $\xi_\phi = 10$ at $N \simeq 2, 2.4, 2.65, 2.8, 2.9$ $e$-folds after the end of inflation (purple, orange, blue, red, green respectively). The black-dotted curve shows a thermal spectrum. }
  \label{fig:spectrum}
\end{figure}

The strong rescattering among fluctuations yields an efficient start to the process of thermalization, by transferring power between particles of different momenta. In Fig.~\ref{fig:spectrum} we show the spectra in field fluctuations $\delta \phi$ and $\delta \chi$ for Case A with $\xi_\phi = 10$. Although the spectra are dominated at early times by increased power in distinct resonance bands, by later times rescattering has flattened out the distributions for both $\delta \phi$ and $\delta \chi$. By $N_{\rm therm} = 2.8$ $e$-folds after the end of inflation, both fields have attained a spectrum consistent with a thermal distribution, $\vert \delta \phi^I_k \vert^2 \propto [ k ( \exp[ k/T ] - 1 )]^{-1}$, at a temperature $T_{\rm reh} \sim {\cal O} (H_{\rm end})$. We find comparable behavior across Cases A and B for $\xi_\phi \geq 1$ \cite{LatticeFollowUp}. 

\begin{figure}[t!]
    \centering
\includegraphics[width=3.2in]{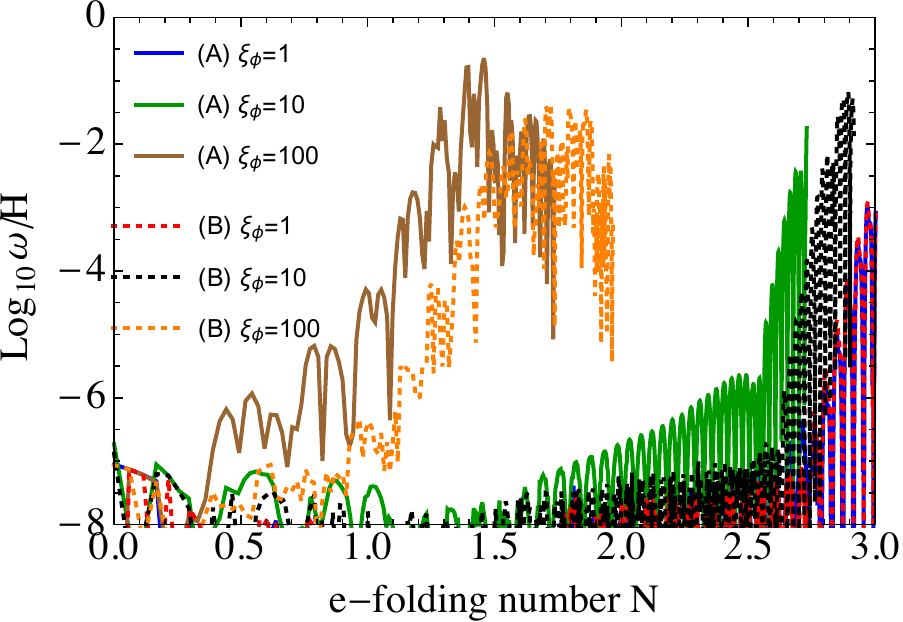}
    \caption{\small The quantity $\vert \omega^I \vert / H$ versus $e$-folds $N$ after the end of inflation for the generic case (A) and the symmetric case (B) and for $\xi_\phi = 1, 10, 100$, where $\omega^I$ is the covariant turn-rate \cite{KMSbispectrum}. Each curve is shown up to $N_{\rm ad} = {\rm min} [ N_{\rm bg}, N_{\rm therm} ]$.
    }
    \label{fig:SFA}
\end{figure}

The rapid thermalization means that the system reaches the adiabatic limit soon after the end of inflation. We denote $N_{\rm ad} = {\rm min} [ N_{\rm bg} , N_{\rm therm} ]$, where $N_{\rm bg}$ is the time by which super-Hubble coherence of the inflaton condensate is lost, indicated by $\phi_{\rm rms} > \langle \phi \rangle$. Any significant turning of the system within the field space between the end of inflation and $N_{\rm ad}$ could amplify non-Gaussianities and isocurvature perturbations, thereby threatening the close agreement between predictions in these models and measurements of the CMB \cite{Elliston:2011dr,Elliston:2014zea,Meyers:2013gua,Turzynski:2014tza}. In Fig.~\ref{fig:SFA}, we plot $\omega / H$ across cases of interest, where $\omega = \vert \omega^I \vert$ is the covariant turn-rate \cite{turnratenote}. Even as the Hubble rate falls over time, we nonetheless find $\omega / H < 0.1$ through $N_{\rm ad}$, indicating minimal turning of the system within field space. 

Our late-time results were unchanged as we varied the initial UV suppression scale $k_{\rm UV} = b H_{\rm end}$ between $b = 25$, $50$, and $100$, and the number of grid-points between $128^3$, $256^3$ and $512^3$.  We discuss this and related numerical convergence tests in Ref.~\cite{LatticeFollowUp}.

{\it Conclusions}. Multifield models of inflation with nonminimal couplings generically yield predictions for primordial observables in close agreement with the latest observations, deriving from the strong single-field attractor behavior of these models \cite{KSprl,SSKisocurv,Preh1}. Throughout the cases we have examined and across parameter space, we find that this single-field attractor behavior remains robust until the system reaches the adiabatic limit after inflation, with no significant turning in field space even in the midst of strongly nonlinear dynamics.

Preheating in this class of models is efficient, draining the energy density from the inflaton condensate within $N_{\rm bg} \lesssim 1.5$ $e$-folds in the limit of strong couplings, $\xi_I \sim 100$. The system typically reaches a radiation-dominated equation of state within $N_{\rm rad} \lesssim 2.5$, while rescattering yields a rapid onset of thermalization within $N_{\rm therm} \lesssim 3$, thereby fulfilling several of the most critical requirements of the reheating phase. We defer to future work such questions as possible impact of coupled metric perturbations on the fully nonlinear preheating dynamics, and the coupling of the scalar fields $\phi$ and $\chi$ to Standard Model particles.

{\it Acknowledgements}. 
RN received support from a Clare Booth Luce Undergraduate Research Award, Grant \#9601.  RN and JTG are supported by the National Science Foundation Grant No. PHY-1719652. JvdV and EIS acknowledge support from the Netherlands Organisation for Scientific Research (NWO). RN, JvdV, and JTG would also like to thank the MIT Center for Theoretical Physics for its warm and generous hospitality. Portions of this work were conducted in MIT's Center for Theoretical Physics and supported in part by the U.S. Department of Energy under Contract No.~DE-SC0012567.

%\bibliography{LatticePreheating}

\begin{thebibliography}{45}
\bibitem [{tur()}]{turnratenote}%
  \BibitemOpen
  \href@noop {} {}\bibinfo {note} {{The covariant turn-rate $\omega^I \equiv
  {\cal D}_t \hat{\sigma}^I$, where $\hat{\sigma}^I \equiv \dot{\varphi}^I /
  \dot{\sigma}$, ${\cal D}_t A^I = \dot{\varphi}^J {\cal D}_J A^I$ is the
  covariant directional derivative, and $\dot{\sigma}^2 \equiv {\cal G}_{IJ}
  \dot{\varphi}^I \dot{\varphi}^J$ \cite{KMSbispectrum}. One may show that
  $\omega = \vert \omega^I \vert = [ V_{, K} V_{, L} (\dot{\sigma}^2 {\cal
  G}^{KL} - \dot{\varphi}^K \dot{\varphi}^L ) ]^{1/2} / \dot{\sigma}^2$. To
  calculate $\omega$ during preheating, we evaluate the spatially homogeneous
  fields via spatial averages on the lattice, $\{ \varphi^I, \dot{\varphi}^I \}
  \rightarrow \{ \langle \phi^I \rangle, \langle \dot{\phi}^I \rangle \}$, and
  replace the term $\dot{\sigma}^2$ in the denominator with its time average
  over each oscillation of the inflaton condensate.}}\BibitemShut {Stop}%
\end{thebibliography}


\end{document}